\documentclass[prd,aps,twocolumn,showpacs,floats,floatfix]{revtex4}
\usepackage[dvips]{graphicx}

\begin{document}

\title{ Tidal deformations of compact stars with crystalline quark matter }

\author{S.~Y. Lau\footnote{Email address: sylau@phy.cuhk.edu.hk}, P.~T. Leung\footnote{Email 
address: ptleung@phy.cuhk.edu.hk}, and L.-M. Lin\footnote{Email address: lmlin@phy.cuhk.edu.hk}  }
\affiliation{Department of Physics and Institute of Theoretical Physics,
The Chinese University of Hong Kong, Hong Kong, China}

\date{\today}

\begin{abstract}
We study the tidal deformability of bare quark stars and hybrid compact stars composed of a quark matter core
in general relativity, assuming that the deconfined quark matter exists in a crystalline color superconducting phase. 
We find that taking the elastic property of crystalline quark matter into account in the calculation of the tidal
deformability can break the universal I-Love relation discovered for fluid compact stars, which connects
the moment of inertia and tidal deformability.
Our result suggests that measurements of the moment of inertia and tidal deformability can in
principle be used to test the existence of solid quark stars, despite our ignorance of the high density 
equation of state (EOS). 
Assuming that the moment of inertia can be measured to 10\% level, one can then distinguish a 1.4 (1) $M_\odot$
solid quark star described by our quark matter EOS model with a gap parameter $\Delta=25$ MeV from a fluid compact star if the tidal deformability can be measured to about 10\% (45\%) level. 
On the other hand, we find that the nuclear matter fluid envelope of a hybrid star can screen out the effect of 
the solid core significantly so that the resulting I-Love relation for hybrid stars still agrees with the universal relation for fluid stars to about 1\% level. 
\end{abstract}


\maketitle

\paragraph*{Introduction.}

The possibility that deconfined quark matter may exist in the ultra-high density cores of compact stars has been of great interest since it was first proposed a few decades ago \cite{Ivanenko65,Itoh70,Baym76_p241}. 
Our current understanding of the QCD phase diagram also suggests that deconfined quarks at the low-temperature and high-density regime can form a condensate of Cooper pairs driven by the BCS mechanism due to the existence of
attractive channels of quark-quark interactions and become color superconducting \cite{Alford98_p247,Rapp98_p81,
Alford99_p443,Alford03_p074024} (see \cite{Alford08_p1455} for a review). 
Soon after their births in supernova explosions, the temperature of compact stars drops quickly below $10^{11}$ K (equivalent to about 10 MeV), the typical transition temperature expected for color superconductivity, and hence it is believed that deconfined quark matter (if exists) inside compact stars can be in a color-superconducting phase.

At the core of compact stars, it may be energetically favorable for quark matter to form an inhomogeneous condensate resulting in a crystalline color-superconducting (CCS) phase \cite{Alford01_p074016,
Casalbuoni05_p89,Mannarelli06_p114012,Rajagopal06_p094019,Casalbuoni06_p350} (see \cite{Anglani14} for a recent review). 
If our current understanding of QCD in the high-density (but still nonperturbative) regime is correct, it is then
possible that hybrid stars featuring a nuclear-matter envelope on top of a CCS quark matter core can exist
\cite{Ippolito08_p023004}. On the other hand, bare solid quark stars composed of CCS quark matter could also be possible if deconfined quark matter is the true ground state of matter \cite{Witten84}.

One of the special properties of the CCS quark matter is that it is extremely rigid. The shear modulus of this crystalline phase can be at least a factor of 20 to 1000 larger than that in traditional neutron star crusts \cite{Mannarelli07_p074026}.
The extreme rigidity of the CCS quark matter may produce detectable gravitational wave signals by sustaining large nonaxisymmetric distortions of the solid core of compact stars \cite{Lin07,Haskell07_p231101,Knippel09_p083007} or exciting the torsional oscillations of the solid core \cite{Lin13}. 
With the groundbreaking detections of gravitational wave signals from binary black holes \cite{LIGO_GW150914,LIGO_GW151226}, we have now entered the era of gravitational wave astronomy. Gravitational waves emitted from binary neutron stars are also expected to be detectable by advanced LIGO and Virgo soon 
in the future. Studying the unique gravitational wave signatures of CCS quark matter is a promising direction to test the existence of this phase of matter.

In this paper, we study the tidal deformability of compact stars with elastic CCS quark matter. The tidal deformability $\lambda$ measures the deformation of a star due to the tidal field ${\cal E}_{ij}$
created by a companion star and is defined by $Q_{ij} \equiv -\lambda {\cal E}_{ij}$, where $Q_{ij}$ is the traceless quadrupole moment tensor of the star.
The tidal deformability $\lambda$ is an important parameter in the study of the gravitational wave signals emitted from compact star binaries. In particular, the parameter characterizes the effects of the internal stellar structure on the gravitational wave phase, and its measurability with gravitational wave 
observations has been studied (see, e.g., \cite{Flanagan_p021502,Damour_p123007,Favata_p101101,Kagi_p021303,
DelPozzo_p071101,Read_p044042,Hotokezaka_p064082}). 
In recent years, the tidal deformability of compact stars has been gaining a lot of attention because of the discovery of the universal I-Love-Q relations connecting the moment of inertia $I$, the tidal deformability $\lambda$ (also called the Love number), and the spin-induced quadrupole moment $Q$ of fluid compact stars \cite{Yagi_p365,Yagi_p023009} (see \cite{Yagi_review} for a recent review). In contrast to the mass-radius relation of compact stars, which depends sensitively on the equation of state (EOS), the I-Love-Q relations are said to be universal because they are approximately EOS-independent to about 1\% level. We refer the reader to \cite{Yagi_p365,Yagi_p023009,Yagi_review} for a detailed discussion of the relevance of these universal relations to astrophysics, gravitational-wave, and fundamental physics.

\begin{figure}
\centering
\includegraphics*[width=9cm]{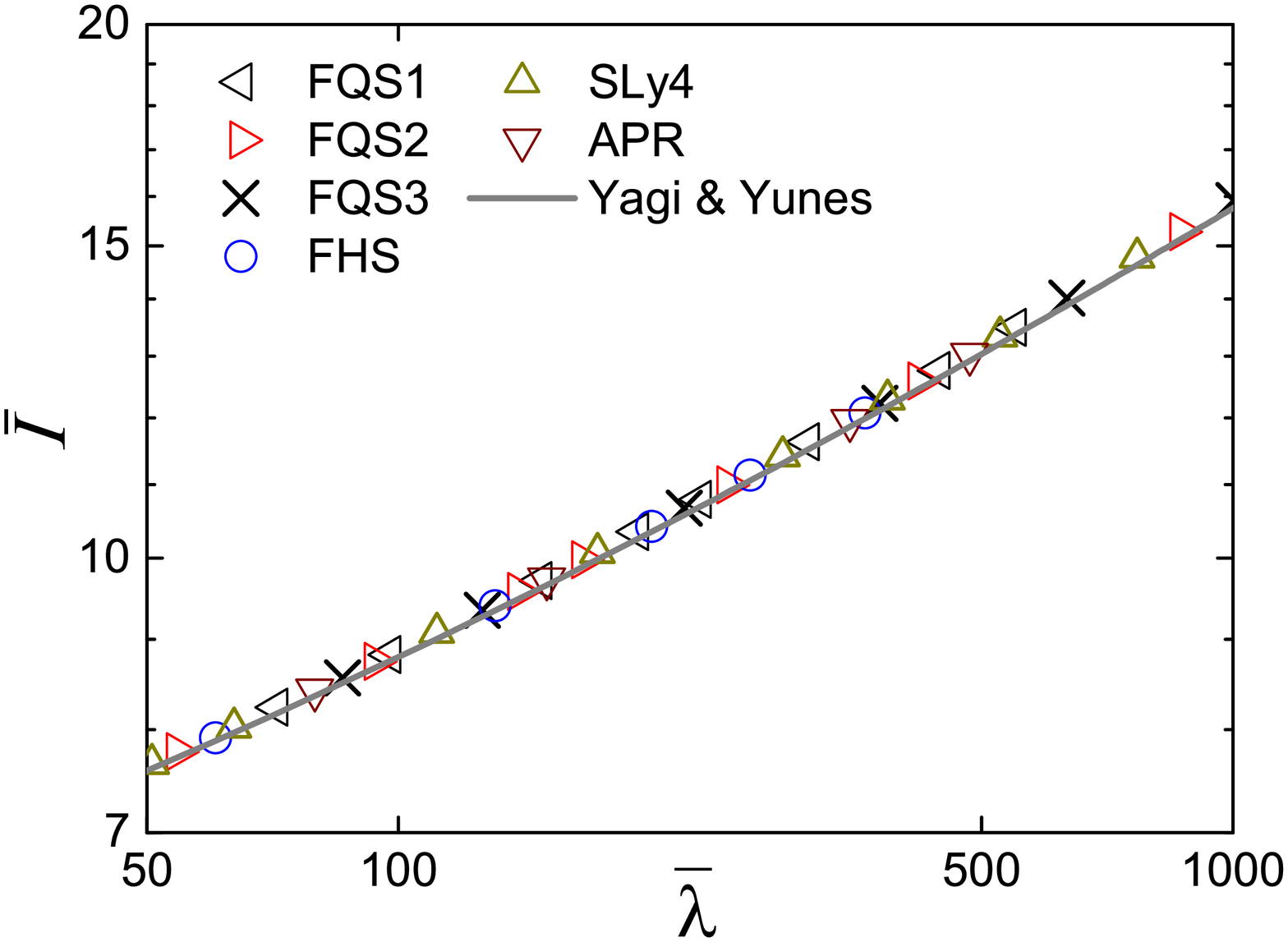}
\caption{Universal I-Love relation for fluid compact stars: $\bar I$ is plotted against $\bar \lambda$ for different stellar models. FQS1, FQS2, and FQS3 represent bare fluid quark stars with different EOS parameters. 
FHS is a hybrid star model composed of a nuclear matter envelope on top of a fluid quark matter core. Traditional neutron stars modeled by the SLy4 and APR EOS models are also plotted. The solid line is a fitting curve for the universal relation proposed by Yagi and Yunes \cite{Yagi_p365,Yagi_p023009}. }
\label{fig:I_Love_Fluid}
\end{figure}

As an illustration, we use a few different EOS models to reproduce in Fig.~\ref{fig:I_Love_Fluid} 
the I-Love relation for fluid stars, which concerns the dimensionless quantities ${\bar I}\equiv I/M^3$ and ${\bar \lambda} \equiv \lambda/M^5$ (in geometric units where $G=c=1$), where $M$ is the gravitational mass. 
Models FQS1, FQS2, and FQS3 represent bare fluid quark star models described by the phenomenological quark-matter EOS defined below in Eq.~(\ref{eq:quark_eos}) with different parameters. FHS is a hybrid star model with a nuclear matter envelope on top of a fluid quark matter core (see below for more details). In the figure, traditional neutron stars described by the nuclear matter EOS models SLy4 \cite{SLy4_eos} and APR \cite{APR_eos} are also presented.
The solid line is a fitting curve suggested by Yagi and Yunes \cite{Yagi_p365,Yagi_p023009}. 
Figure~\ref{fig:I_Love_Fluid} shows clearly that the relation between $\bar I$ and $\bar \lambda$ (the I-Love relation) is insensitive to the underlying EOS models. The internal structure of bare quark stars and hybrid stars are very different from traditional neutron stars because they either have a finite non-zero density at the surface or a density discontinuity in the interior. It is thus quite surprising that these different stellar models can establish the same universal I-Love relation.
In the following, we shall investigate how the contribution of the elasticity of crystalline quark matter in the calculation of $\lambda$ would affect the I-Love relations for bare solid quark stars and hybrid stars with a solid quark core.

\paragraph*{Formulation.}

Here we briefly outline the formulation of our study. 
The problem is to determine tidal quadrupolar deformations of an elastic star perturbatively.
The unperturbed background star is assumed to be in an unstrained state and can be treated as a perfect fluid 
star obeying the Tolman-Oppenheimer-Volkoff (TOV) equation. The tidal deformation of the star is considered perturbatively by linearizing the Einstein equations and the conservation equation for the matter field:
\begin{equation}
\delta G_{\alpha\beta} = 8\pi \delta T_{\alpha\beta} ,
\end{equation}
\begin{equation}
\delta \left( \nabla^{\alpha} T_{\alpha\beta} \right) = 0 ,
\end{equation}
where $G_{\alpha\beta}$ is the Einstein tensor and $T_{\alpha\beta}$ is the stress-energy tensor.
$\nabla^{\alpha}$ and $\delta$ denote the covariant derivative and Eulerian perturbation, respectively. 
The elasticity of the matter contributes through the shear part $T^{\rm shear}_{\alpha\beta}$ of the full 
stress-energy tensor
\begin{equation}
T_{\alpha\beta} = T^{\rm bulk}_{\alpha\beta} + T^{\rm shear}_{\alpha\beta} ,
\end{equation}
where the bulk part $T^{\rm bulk}_{\alpha\beta}$ is assumed to take the standard perfect-fluid form. As 
the background star is assumed to be in an unstrained state, $T^{\rm shear}_{\alpha\beta}$ in fact 
contributes only at the perturbation level through \cite{Finn90,Penner_103006,Kruger_063009}
\begin{equation}
\delta T^{\rm shear}_{\alpha\beta} = - 2 \mu \delta \Sigma_{\alpha\beta} , 
\end{equation}
where $\mu$ is the shear modulus and $\delta \Sigma_{\alpha\beta}$ is the perturbed shear tensor generated by 
a displacement field due to the quadrupolar deformation. 

In the case of a bare solid quark star, the problem can be cast into a system of six coupled first-order
ordinary differential equations for the metric and matter perturbation variables. 
The equations are solved by noting that there are three regular solutions at the center, together with the vanishing of the radial and tangential stresses at the stellar surface. 
The interior solution is then determined up to an arbitrary constant. Once the interior problem has been solved, the remaining procedure to determine the tidal deformability $\lambda$ is the same as that for bare fluid quark stars.
By using the determined metric function $g_{tt}$ and comparing with the response of a static 
star to an external quadrupolar tidal field ${\cal E}_{ij}$ in the far-field limit: 
\begin{equation}
-{ 1 + g_{tt} \over 2}  =  -{ M \over r} - {3 Q_{ij}\over 2 r^3} \left( {x^i x^j\over r^2} - {1\over 3
}\delta^{ij}
\right) + {1\over 2} {\cal E}_{ij} x^i x^j , 
\end{equation}
where $Q_{ij}$ is the quadrupole moment, the tidal deformability $\lambda$ of 
the star is then determined by the relation $Q_{ij}=-\lambda {\cal E}_{ij}$
(see, e.g., \cite{Postnikov_024016} for a detailed discussion).

For the case of a hybrid star, the perturbation equations describing the solid core are the same as those for 
a solid quark star with the same regular solutions at the center. At the solid-fluid interface where the density and shear modulus are discontinuous, we impose the continuities of the metric variables and the radial and tangential stresses across the interface. In the fluid layer on top of the solid core, the shear modulus $\mu=0$ and the perturbed bulk part $\delta T^{\rm bulk}_{\alpha\beta}$ is the only contribution to the stress-energy tensor. One only needs to solve a second-order differential equation for the metric perturbation $\delta g_{tt}$ 
in order to determine $\lambda$.

We also note that the quadrupolar oscillations of neutron stars taking into account the elasticity of the 
neutron star crust have been studied by Finn \cite{Finn90} and Kr\"{u}ger {\it et al.} \cite{Kruger_063009}. 
We have checked that our final set of perturbation equations for crystalline matter agrees with their relevant equations in the zero-frequency limit.


\paragraph*{Microphysics Input.}

In order to construct equilibrium stellar models, we employ the phenomenological quark-matter EOS model of Alford {\it et al.} \cite{Alford05_p969} for the quark matter. The model is defined by the thermodynamic potential
\begin{equation}
\Omega_{\rm QM} = - { 3 \over 4 \pi^2} a_4 \mu_q^4
+ {3 \over 4 \pi^2} a_2 \mu_q^2 + B_{\rm eff} ,
\label{eq:quark_eos}
\end{equation}
where $\mu_q$ is the quark chemical potential. The parameter $a_4\ (\le 1)$ is used to model nonperturbative
QCD corrections. The reasonable value for the dimensionless parameter $a_4$ is expected to be of order 0.7 \cite{Alford05_p969}. The parameter $a_2$ is used to model the effects of quark masses and pairing. $B_{\rm eff}$ is an effective bag constant. The energy density $\rho$ and pressure $P$ for quark matter can be calculated from $\Omega_{\rm QM}$ using standard thermodynamic relations: $\rho = \Omega_{\rm QM} + n_q \mu_q$ and 
$P=-\Omega_{\rm QM}$, where $n_q = - \partial \Omega_{\rm QM}/\partial \mu_q$ is the quark number density.  
Using the relation between $\rho$ and $P$, together with the TOV equation, allows us to construct static bare quark stars. In the following, we fix the parameters $a_4=0.8$, $a_2^{1/2}=100$ MeV, and $B_{\rm eff}^{1/4} = 160$ MeV.

As we have discussed, the shear modulus of CCS quark matter is a key parameter in the calculation of the tidal deformability of compact stars featuring crystalline quark matter.
It has been calculated by Mannarelli {\it et al.} \cite{Mannarelli07_p074026} and is given by
\begin{equation}
\mu = 2.47 {\rm \ MeV/fm^3} \left( {\Delta\over 10{\ \rm MeV} } \right)^2
\left( {\mu_q \over 400 {\ \rm MeV} }\right)^2 ,
\label{eq:mu}
\end{equation}
where the gap parameter $\Delta$ is expected to be in the range $5\ {\rm MeV} \lesssim \Delta \lesssim
25\ {\rm MeV}$ \cite{Mannarelli07_p074026}.
The rigidity of the crystalline matter is characterized by the shear modulus, and hence the gap parameter $\Delta$. For a given stellar density profile, a larger $\Delta$ corresponds to a more rigid crystalline quark matter as suggested by Eq.~(\ref{eq:mu}). We shall regard $\Delta$ as a free parameter in the following investigation.

For hybrid stars featuring a solid quark core, we use the same quark-matter EOS model to describe the quark core. In the nuclear-matter envelope, we employ the APR EOS model \cite{APR_eos}. The phase transition from nuclear matter to quark matter is implemented using a Maxwell construction (see \cite{Alford05_p969,Lin13} for more details).


\paragraph*{Results.}

\begin{figure}
\centering
\includegraphics*[width=9cm]{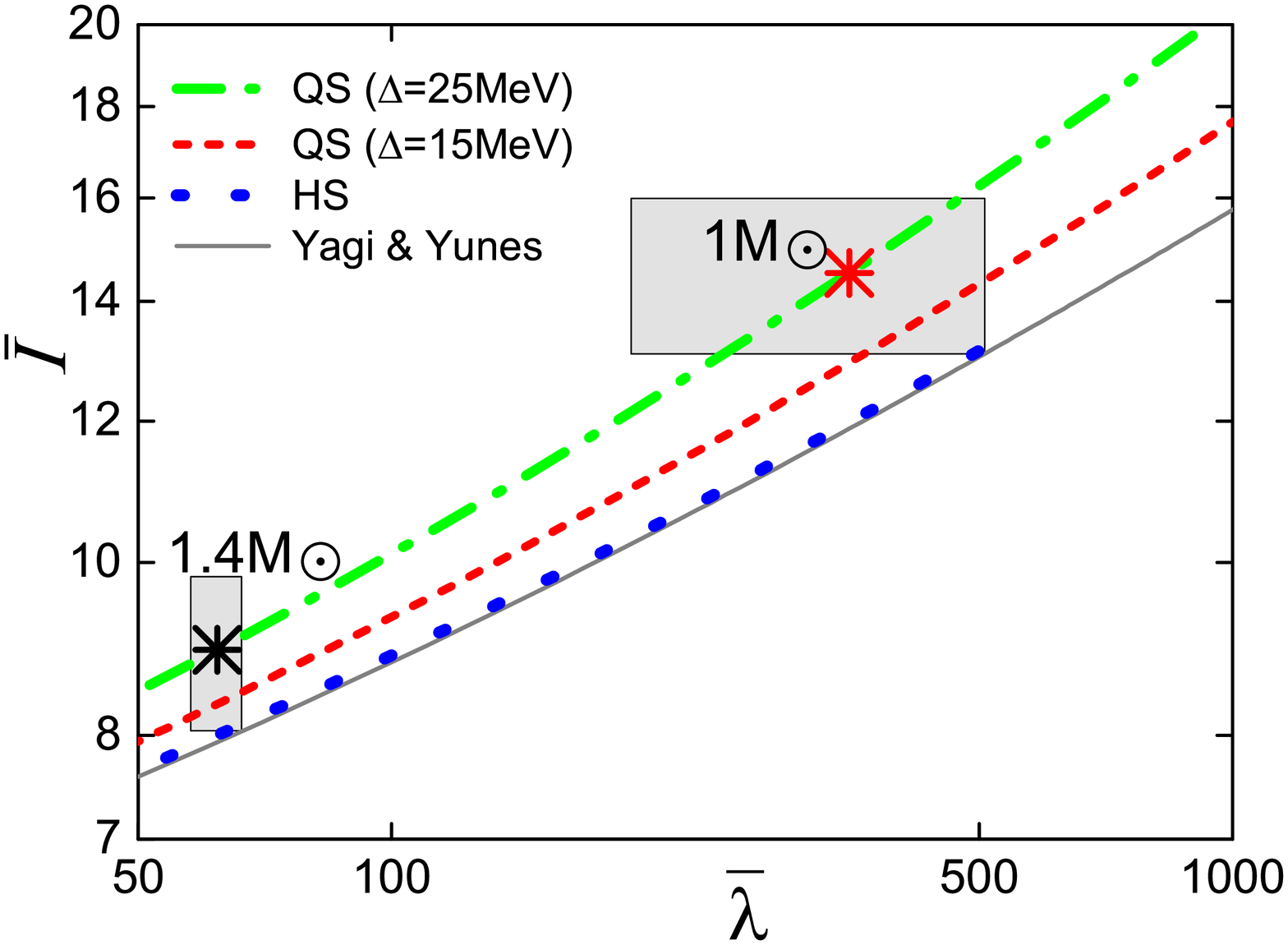}
\caption{ $\bar I$ is plotted against $\bar \lambda$ for bare solid quark stars (QS) and hybrid stars (HS) with 
a solid quark matter core. Two values of the gap parameter $\Delta=15$ and 25 MeV are used for QS models, while $\Delta=25$ MeV is used for HS models. The universal I-Love relation (solid line) for fluid stellar models proposed by Yagi and Yunes \cite{Yagi_p365,Yagi_p023009} is also plotted for comparison. The shaded area around the 1 and 1.4 $M_\odot$ solid quark stars marked in the figure represents an observation error box in the I-Love plane in order to distinguish these solid stars from fluid compact stars.
}
\label{fig:I_Love_Solid}
\end{figure}


As we have seen in Fig.~\ref{fig:I_Love_Fluid}, both bare quark-star and hybrid-star models have essentially the same universal I-Love relation as traditional neutron stars, assuming that the deconfined quark matter inside the stars is in a fluid state. We are now ready to use the formulation outlined above to study the effect of crystalline quark matter on the I-Love relation by taking into account the elasticity of crystalline matter in the
calculation of the tidal deformability.
Figure~\ref{fig:I_Love_Solid} presents our main result where the I-Love relations for bare solid quark stars
(QS) and hybrid stars (HS) with a solid core are shown, together with the fitting curve of Yagi and Yunes 
(solid line) \cite{Yagi_p365,Yagi_p023009} for fluid compact stars. 
Two different values of the gap parameter $\Delta$ (15 MeV and 25 MeV) are considered for the bare quark star models. It is seen clearly that the I-Love relation of bare solid quark stars depends on the value $\Delta$ and deviates significantly from the universal relation. In particular, the deviation increases with the value of $\Delta$, and hence the rigidity of the quark matter. 
In the figure, we also mark the positions of two quark star models with masses $1 M_\odot$ and $1.4 M_\odot$ on the I-Love relation for the case $\Delta=25$ MeV for reference. We note that the tidal deformability of a solid quark star decreases with increasing $\Delta$. This is a predictable result because increasing $\Delta$, and hence the shear modulus for a given background model, means that the star is more difficult to be deformed tidally. 
In particular, the dimensionless quantity $\bar \lambda$ of a $1.4 M_\odot$ quark star model decreases from 106.1 to 63.5 as the quark matter changes from a fluid state to a solid state with $\Delta=25$ MeV. 
For comparison, including the elasticity of the neutron star crust only leads to a tiny change in the tidal 
deformability of traditional neutron stars \cite{Penner_103006} because the shear modulus of neutron star crust is 
much smaller than that of crystalline quark matter, and the mass fraction of the thin crust is too small to 
produce a significant effect.

\begin{figure}
\centering
\includegraphics*[width=9cm]{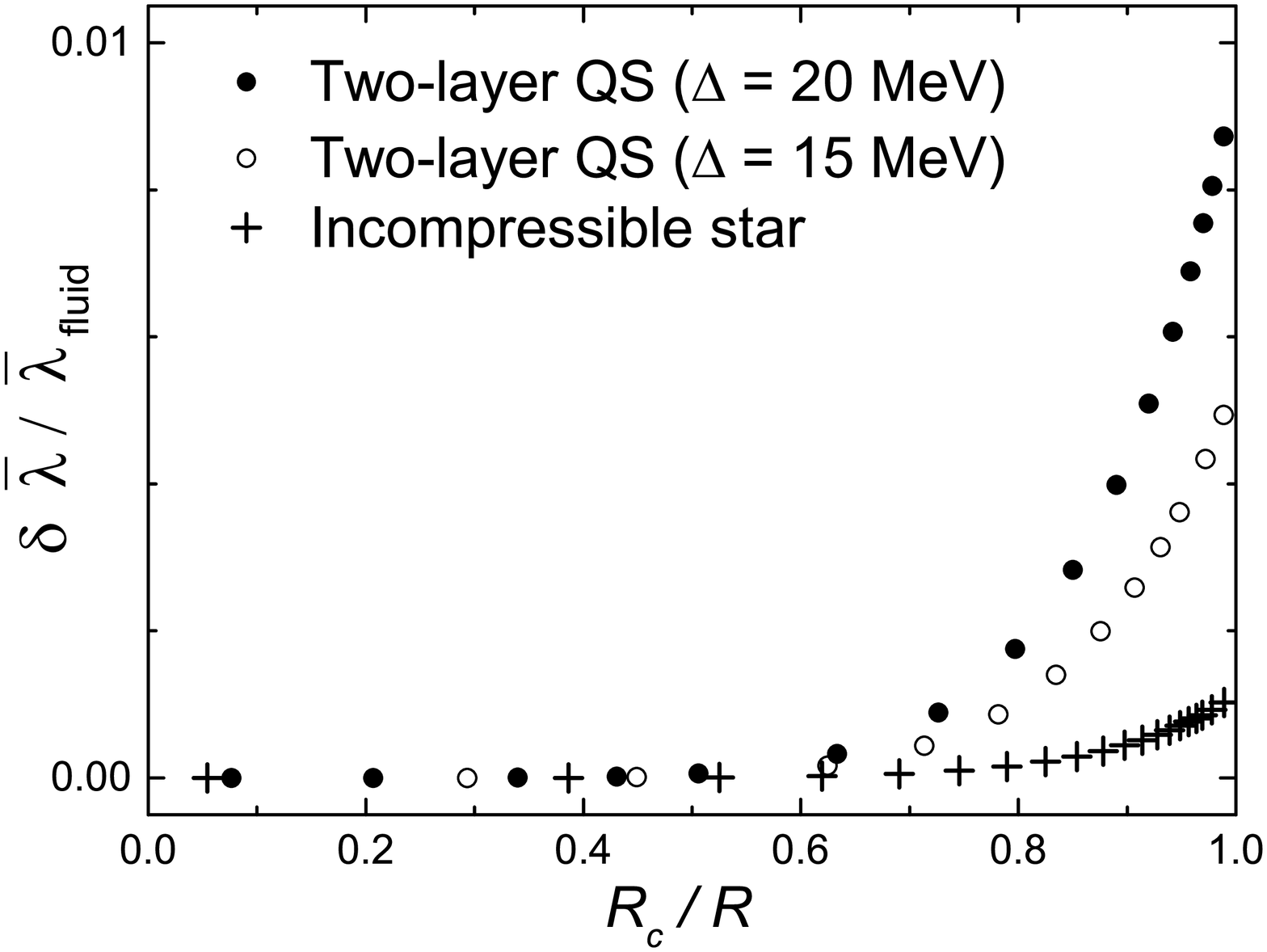}
\caption{Screening effect: The fractional difference $\delta\bar\lambda / {\bar \lambda}_{\rm fluid}$ as defined in the text is plotted against $R_c/R$ for two-layer quark star and incompressible models. Both models are assumed
to have a fluid envelope on top of a solid core. }
\label{fig:Screen}
\end{figure}

An implication of our result is that independent measurements of $\bar I$ and $\bar \lambda$ can in principle be used to test the existence of bare solid quark stars, despite our ignorance of the high-density EOS model, by
considering the deviation of the measured values from the universal relation of Yagi and Yunes (solid line
in Fig.~\ref{fig:I_Love_Solid}). 
For example, let us assume the existence of a $1.4 M_\odot$ bare solid quark star composed of CCS quark matter with $\Delta=25$ MeV in nature. We note that the evidence for a solid quark star in this case is strong if both $\bar I$ and $\bar \lambda$ can be measured to about 10\% accuracy. The shaded area around the $1.4 M_\odot$ solid quark star marked in Fig.~\ref{fig:I_Love_Solid} represents the observation error box (see \cite{Yagi_p365,Yagi_p023009} for the construction of a similar observation error box in the content of distinguishing a modified gravity theory from general relativity based on the universal relation). 
It should be noted that future observations may indeed be able to measure the moment of inertia in double pulsar systems to about 10\% accuracy \cite{Lattimer_Schutz}.  
We also note that the requirement on the measurement accuracy is less stringent for lower mass stars. 
Let us consider the $1 M_\odot$ solid quark star marked in Fig.~\ref{fig:I_Love_Solid}. Assuming that $\bar I$ of this star can still be measured to 10\% accuracy, the corresponding accuracy level that is required for $\bar \lambda$ in order to distinguish this solid quark star from a traditional neutron star (or fluid quark star) is about 45\% instead of 10\%. The shaded area around this star in Fig.~\ref{fig:I_Love_Solid} is the corresponding error box.


In Fig.~\ref{fig:I_Love_Solid} the I-Love relation of hybrid star (HS) models is obtained by using the gap
parameter $\Delta=25$ MeV. Having seen that the I-Love relation of bare solid quark stars deviates significantly from the universal relation of fluid stellar models, one might expect that the solid core inside hybrid star models, despite being enclosed by a nuclear-matter fluid envelope, should also produce a comparable effect. However, Figure~\ref{fig:I_Love_Solid} shows that this is not the case. Although we have maximized the rigidity of the solid core by employing the theoretical upper bound for the gap parameter $\Delta=25$ MeV, the I-Love relation of hybrid stars still agrees very well (about 1\% level) with that of the fluid star models. 
For a $1.4\ M_\odot$ hybrid star model, the dimensionless quantity $\bar \lambda$ changes only from 212.6 to 211.4 when the quark matter core of the star changes from a fluid state to a solid state. In contrast to the case in bare solid quark stars, the effect of the solid core in hybrid stars is unexpectedly small. This suggests that the fluid envelope can somehow ``screen'' out the effect of the solid core so that the presence of the solid core cannot be revealed easily by observing the tidal deformation of hybrid stars.

To further illustrate the screening effect, we have also considered a two-layer bare quark star model and an incompressible uniform density model. Both models are assumed to have a fluid envelope on top of a solid core.
In contrast to the hybrid star models where the radius $R_c$ of the solid core is fixed by the transition pressure of the underlying EOS, we can now regard $R_c$ as a parameter and study the screening effect by varying its value from 0 to $R$, where $R$ is the radius of the star.  

In Fig.~\ref{fig:Screen} we plot the fractional difference $\delta{\bar \lambda} / {\bar \lambda}_{\rm fluid}$ against the ratio $R_c/R$. Here ${\bar \lambda}_{\rm fluid}$ is the scaled tidal deformability for a pure fluid star (i.e., when the radius of the solid core $R_c=0$) and $\delta {\bar \lambda} \equiv {\bar \lambda}_{\rm fluid}-{\bar \lambda}(R_c)$, where ${\bar \lambda}(R_c)$ is the scaled tidal deformability of a star model with 
a solid core of radius $R_c$. For the quark star model, two different values of the gap parameter 15 and 20 MeV are used to calculate the shear modulus of the CCS quark matter. On the other hand, a constant shear modulus is 
used for the incompressible model.  
Figure~\ref{fig:Screen} shows that the screening effect is almost complete ($\delta\bar\lambda \approx 0$) even for a solid core as large as $R_c \approx 0.7 R$ for both stellar models. For a fixed value of $R_c$, it is noted that $\delta \bar\lambda$ increases with the gap parameter $\Delta$, and hence the rigidity of the solid core, in the two-layer quark star model as one would expect. We also note that the screening effect in the incompressible model is much stronger. 
For the incompressible model, we find that $\delta{\bar \lambda} / {\bar \lambda}_{\rm fluid}$ is of the order $10^{-3}$ even when the radius of the solid core is up to $R_c = 0.99 R$. 

Interestingly, this somewhat surprising screening effect has also been found recently by Beuthe \cite{Beuthe} in his study of the tidal deformation of icy satellites of the solar system, such as Europa and Titan. In particular, 
Beuthe demonstrated analytically the existence of a complete screening effect ($\delta\bar\lambda=0$) in Newtonian
gravity for a two-layer uniform density stellar model, with a surface fluid layer on top of a solid core. In particular, the screening effect is independent of the thickness of the fluid layer. A generalization of the analytical work of \cite{Beuthe} in Newtonian gravity to general relativity would be an interesting future investigation.


\paragraph*{Discussion.}

Besides gravitational wave signals \cite{Lin07,Haskell07_p231101,Knippel09_p083007,Lin13} that we have mentioned before, electromagnetic \cite{Mannarelli_p103014} and neutrino emissions \cite{Anglani_p074005} from compact stars composed of CCS quark matter have also been investigated. However, these signals in general depend sensitively on the underlying EOS model, which is not well understood even for traditional neutron star models. 
In this work, we propose that an observed broken universal I-Love relation from independent measurements of 
even one single pair of $\bar I$ and $\bar \lambda$ would provide us a strong evidence for the existence of solid quark stars, and hence verifying (i) the hypothesis that quark matter could be absolutely stable and (ii) the existence of the CCS phase {\it at the same time}, despite our ignorance of the high-density EOS. 
For instance, assuming that the gap parameter takes the expected maximum value $\Delta =25$ MeV and both $\bar I$ and $\bar \lambda$ can be measured to about 10\% accuracy, we can then distinguish a $1.4 M_\odot$ solid quark star modeled by our quark-matter EOS from a traditional neutron star. Furthermore, the required accuracy on $\bar \lambda$ is reduced to about 45\% if the mass of the solid quark star is $1 M_\odot$. 
On the other hand, if the pair of values agree with the universal relation, it does not necessary rule out the 
CCS phase because this exotic phase of matter can still exist inside hybrid stars which have essentially the same
I-Love relation as traditional neutron stars due to a screening effect.

We end this paper with a few remarks. (1) Our conclusion is based on the assumption that general relativity is 
the correct theory of gravity. Otherwise, a deviation from the universal I-Love relation may in fact be 
a signature of a modified gravity theory \cite{Yagi_p365,Yagi_p023009,Sham_p66,Pani_p024025} instead of 
the CCS quark matter. If this is the case, then one can still hope to distinguish the two situations by making use of an extra observable such as the stellar mass $M$, which is already assumed to be known when constructing the scaled quantities $\bar I$ and $\bar \lambda$.
(2) For a binary system approaching merger, the effects of spin on the tidal deformability can be important and would limit our ability to identify solid quark stars using the universal relations \cite{Pani_p124003}. (3) Nevertheless, according to neutron-star binary simulations (see, e.g., \cite{Sekiguchi}), the internal temperature of the stars during merger can rise up to a few tens of MeV. If this temperature scale also applies to a quark-star binary, depending on the value of the gap parameter, the CCS phase of quark matter may in fact already disappear when the system is close to merger.

\bibliographystyle{prsty}

\end{document}